\documentclass[epsfig,twocolumn,pra,aps,floats]{revtex4}
\usepackage{epsf,psfig,graphicx,latexsym,amsmath}
\def\v{{\rm v}}
\def\deltan{{\delta}}
\def\eff{{\rm eff}}

\def\comment#1{}
\newcommand{\BF}[1]{\mbox{\boldmath $#1$}}
\begin{document}
\title{Reentrant Phenomenon in Quantum Phase Diagram of
Optical Boson Lattice}
\author{H.~Kleinert$^{1}$, S. Schmidt$^{2}$, and A. Pelster$^{3}$}
\affiliation{$^{1}$Institut f\"ur Theoretische Physik,
Freie Universit\"at Berlin, Arnimallee 14, 14195 Berlin, Germany\\
$^{2}$Department of Physics, Yale University, P.O. Box 208120, New Haven, CT 06520-8120, USA\\
$^{3}$Fachbereich Physik, Universit\"at Duisburg-Essen, Universit\"atsstrasse 5, 45117 Essen, Germany}
\begin{abstract}
We calculate the
location of the quantum phase transitions of
a  bose gas trapped in an optical lattice as a function of
effective scattering length $a_{\eff}$ and temperature $T$.
Knowledge of recent high-loop results on the shift of the critical
temperature at weak couplings is used to locate a {\em nose\/} in
the phase diagram above the free Bose-Einstein critical
temperature $T_c^{(0)}$, thus predicting the existence of a
reentrant transition {\em above\/}   $T_c^{(0)}$, where
a condensate should form when {\em increasing\/} $a_{\eff}$.
At zero temperature, the transition
to the normal phase produces
the experimentally observed Mott insulator.
\end{abstract}
\maketitle
Optical lattices offer the intriguing possibility
of investigating properties of
Bose-Einstein condensates (BECs) at
varying effective interaction strengths \cite{Fisher,Jaksch}.
Due to Bloch's theorem,  particles loaded into such lattices
behave in some ways like particles in an ordinary continuous space.
Thus they may form a BEC condensate
in the zero-momentum state
in the same way as in the
continuum.

Because of the lattice structure,
the bosons in an optical lattice
exhibit another phase transition which does not exist
in the continuum \cite{Denteneer}.
It produces a
state in which the atoms are frozen to their individual
optical potential wells on the lattice
and require an activation energy to be moved.
This state
is referred to as a
{\em Mott insulator\/} due to its analogy to a state described
long ago by Mott
for  fermionic systems \cite{Mott}.
For bosons on a lattice, a Mott insulator is generally
expected to exist for
integer filling factors at all temperatures
if the repulsion between the atoms
is larger than $k_BT$ so that
the atoms have no place to escape from
their optical potential wells.

At very low temperature,
the existence of this state
has recently been demonstrated
by determining the critical effective scattering length \cite{Greiner}.
It is generally expected
that at zero temperature the
Mott transition
should coincide with the  superfluid-normal
transition.
At the critical point, the excitation energies of
the bosons acquire a gap which pins
the atoms to their
potential wells.
Expressed differently, the Goldstone modes
of translations have become massive
and the associated phase fluctuations decoherent,
in accordance
with the criterion found in Ref.~\cite{CR}.

Standard treatments of this situation proceed by analytic or numeric studies of the Bose-Hubbard model
\cite{Fisher,Jaksch,HFB1,HFB2}.
The purpose of this paper is to simplify the model by focusing the attention upon systems with low-density
where physics takes place in the lowest energy-band. There simple field-theoretic techniques can be applied
to calculate the whole quantum phase diagram
for the superfluid-normal transition.
Our main point will be to include recent high-loop results
near $T_c$ of the weakly coupled system to predict an
interesting characteristic reentrant transition.
In addition  we shall observe
that our calculated superfluid-normal
 transition point at zero temperature
agrees with the experimentally observed
Mott transition point.
\comment{
with summarizing the physical background of these fascinating experiments in optical lattices.
In particular, we highlight the quasifree particle view of the optical boson lattice which is due to the Bloch theorem.
Section {\bf 2)} then presents a general theory for the quantum phase transition of an effectively homogeneous Bose gas
for an arbitrary single-particle dispersion. Then we specialize our result in Section {\bf 3)} for a
free-particle spectrum, yielding a quantum phase diagram with a characteristic reentrant transition. Finally, Section {\bf 4)}
presents a calculation of the transition line for a band structure of an optical lattice.}\\

{\bf 1)}  If bosons of mass $M$  are trapped in a three-dimensional cubic periodic potential $V({\bf x})$
of lattice vectors ${\BF \deltan}$, i.e.,
$V({\bf x})=V_0 \sum _{i=1}^3 \sin^2 (q_i x_i)$ with $q_i=\pi/\deltan$,
the wave vector ${\bf q}$ defines an energy scale
$E_r={\hbar ^2 {\bf q}^2}/{2M}$.
If the individual potential wells are deep, i.e.,
$V_0 \gg E_r$, the single particle Wannier functions $w({\bf x})$
in the nearly harmonic wells are given by oscillator
ground-state wave functions at the lattice sites ${\BF \deltan}$
with size $A_0= \sqrt{\hbar /M \omega _0}$ and energy
$\hbar  \omega _0\approx2E_r\left(V_0/E_r\right)^{1/2}$. The
lowest energy band arising
due to Bloch's theorem
reads $\epsilon_B ({\bf k})= 2  J \sum _{i=1}^3[1-\cos (k_i\deltan)]$ up to a trivial additive constant
Here $J$ follows from the tight-binding approximation
as $J=\int d^3xw({\bf x})[-\hbar ^2{\BF \nabla}^2/2M+V({\bf x})]w({\bf x}+
{\BF \delta})$ and is equal to
$J=  4E_r  (V_0/ E_r)^{3/4}
e^{-2 (V_0/E_r)^{1/2}}/ \sqrt{\pi}$ \cite{Zwerger}.
%
%\begin{equation}
%J=   \frac{4}{ \sqrt{\pi} }  \, E_r\left(\frac{V_0}{E_r}\right) ^{3/4}
%\exp\left[-2\left(\frac{V_0}{E_r}\right) ^{1/2} \right]\,.
%\label{WIDTH}
%\end{equation}
%
Due to the low-density of the system, a repulsive potential
$V({\bf x})=g\delta^{(3)}({\bf x})$ with the coupling constant
$g=4\pi \hbar ^2 a/M$
approximates well all relevant
spherically symmetric short-range two-particle interactions, where
$a$ is the $s$-wave scattering length.
In an optical lattice, this gives rise to an effective repulsive
$\delta$-function interaction with strength
$U\equiv g\int d^3 x\, w^4({\bf x})= (a/A_0)
2 \hbar  \omega _0 / \sqrt{2\pi}= (2\pi a/ \lambda )\sqrt{8/\pi}E_r(V_0/E_r)^{3/4}$ \cite{Jaksch,Zwerger}.
The importance of the interactions
between the particles
in the periodic traps
is measured by the ratio $\gamma\equiv U/J$
between interaction  energy $U=g_{\eff}n$ with $g_{\eff}=4\pi \hbar ^2 a_{\eff}/M_{\eff}$
and kinetic energy $J = \hbar^2 n^{2/3}/2M_{\eff}$, where $n$
is the particle density ($=f/\deltan^3$ with filling factor $f$).
This leads to $\gamma = 8 \pi a_{\eff}\,n^{1/3}$.

The experimental optical lattice of Ref.~\cite{Greiner} is made of
laser beams with wavelength $\lambda$ = $2 \delta$ = 852 nm and
contains about $2\times 10^5$ atoms ${}^{87}$Rb with
$a\approx4.76$ nm \cite{sl}. Its energy scale is
%$E_r\approx3$\,kHz $\mathop{\hat=}   23$ nK,
{$E_r\approx \hbar\times
20$\,kHz $\mathop{\approx} k_B\times  150$ nK} and $V_0/E_r$ is raised from
12 to 22. In this range, $J/E_r$ drops from 0.014 to 0.002,
$U/E_r$ increases from 0.36 to 0.57, and $\hbar  \omega _0/E_r$
increases from 0.36 to 0.57. Expanding the
small-${\bf k}$ behavior of the band energy
$\epsilon_B({\bf k})$
as $\hbar ^2{\bf k}^2/2M_{\eff}+\dots$,
the band width $4J$
defines an effective mass $M_{\eff}$ of the particles $M_{\eff}=\hbar
^2/2J\deltan^2$. In a typical BEC with $a_{\eff}$ of the order of
\AA~and particle distances of a few thousands \AA, this ratio is
extremely small. For the particles tightly bound in an optical
lattice, however, $a_{\rm eff}\,n^{1/3}$ can be made quite large.
In the experiment of Ref.~\cite{Greiner}, for temperatures near zero we
have $\gamma \approx 0.0248 \exp(2 \sqrt{V_0/E_r})$, so that the
increase of the potential depth $V_0/E_r$ from 12 to 22
raises $a_{\eff} n^{1/3}$ from 1 to 11.7.

For increasing temperatures, we expect the critical
$a_{\eff} n^{1/3}$  to decrease until it hits zero
as $T$ reaches
roughly
the free
BEC critical temperature
$T_c^{(0)}=2\pi \hbar ^2 [n/\zeta(3/2)]^{2/3}/M_{\eff}k_B$
with $\zeta(3/2)\approx 2.6124$.
In the above experiment where
 $V_0/E_r$ is raised from 12 to 22, the
temperature $T_c^{(0)}$ drops from
14.2
\,nK to 1.93 \,nK, implying that $T_c^{(0)}/E_r$ drops from 0.094
to 0.013. Hence
$J$ and $k_BT$ are much smaller
than $\hbar  \omega _0$, so that we  can ignore all higher bands.

The purpose of this note is to derive
the full temperature dependence
of this transition, thereby predicting a surprising
reentrant phenomenon.\\

{\bf 2)} We begin by considering a $D$-dimensional Bose gas in the dilute limit where the
two-particle $ \delta $-function interaction is dominant. In the grand-canonical ensemble
it is described by the Euclidean action
\vspace{-.6em}
\begin{eqnarray}
{\cal A}[\psi^\ast,\psi] \!\!&=&\!\! \int_0^{\hbar \beta}d\tau \int d^D x
\Big\{ \psi^\ast ({\bf x},\tau)[ \hbar\partial_\tau\! +\! \epsilon
(-i\hbar{\BF \nabla}) \!-\!\mu] \nonumber
 \\&&
\times
\psi({\bf x},\tau)
+\frac{g_{\eff}}{2}\, \psi({\bf x},\tau)^2
\psi^\ast({\bf x},\tau)^2  \Big\} \,,
\\[-1.2em] \nonumber
\end{eqnarray}
where $\mu$ is the chemical potential, and $\beta\equiv 1/k_B T$.
To describe the
phase transitions
in this gas we
calculate its effective energy.
We expand the Bose field $ \psi ({\bf x},\tau)$
around a constant background $\Psi$, i.e.,
$\psi({\bf x},\tau)=\Psi+\delta\psi({\bf x},\tau)$,
and perform the functional integral
for the grand-canonical partition function including only
the harmonic fluctuations in
$\delta\psi ({\bf x},\tau)$. This yields
the one-loop approximation to the effective potential
\vspace{-.6em}
\begin{eqnarray}
\label{loop0}
{\cal V} (\Psi,\Psi^\ast) &=& V \left( -\mu|\Psi|^2+\frac{g_{\eff}}{2}|\Psi|^4 \right) +\frac{\eta}{2}
\sum_{{\bf k}}E({\bf k}) \nonumber \\
&& + \frac{\eta}{\beta}\sum_{{\bf k}} \ln \left[1-e^{-\beta E({\bf k})} \right]\,,
\label{@VEFF}
\\[-1.2em] \nonumber
\end{eqnarray}
with
%quasiparticle energies
%
%\begin{eqnarray}
%\label{loop1}
$E({\bf k})=\sqrt{\{\epsilon({\bf k})-\mu+2g_{\eff}|\Psi|^2\}^2-g_{\eff}^2|\Psi|^4}$
%\,.
%\end{eqnarray}
%
denoting the quasiparticle energies.
An expansion parameter
$ \eta =1$ has been introduced
whose power serves to
count the loop order.
The effective potential (\ref{loop0})  is extremized with respect to the background field $\Psi$, yielding
the condensate density
\vspace{-.6em}
\begin{eqnarray}
\label{bog4}
n_0&=& \Psi^* \Psi = \frac{\mu}{g_{\rm eff}}-\frac{\eta}{V}\sum_{{\bf k}}\frac{2\epsilon({\bf k})+\mu}{\sqrt{\epsilon({\bf k})^2+2\mu
\epsilon({\bf k})}}
\nonumber \\ &&
\times \left(\frac{1}{2} +\frac{1}{e^{\,\beta
\sqrt{\epsilon({\bf k})^2
+2\mu \epsilon({\bf k})}}-1}\right)+{\cal O}(\eta^2)
\\[-1.2em] \nonumber
\end{eqnarray}
and the grand-canonical potential
\vspace{-.6em}
\begin{eqnarray}
\label{bog6}
&&\frac{\Omega (\mu,T)}{V} = -\frac{\mu^2}{2g}+\frac{\eta}{2V}\sum_{{\bf k}}\sqrt{\epsilon({\bf k})^2+2\mu \epsilon({\bf k})}
\nonumber \\
&& + \frac{\eta}{\beta V}\sum_{{\bf k}}\ln \left(1-e^{-\beta \sqrt{\epsilon({\bf k})^2+2\mu \epsilon({\bf k})}}\,\,\right)
+{\cal O}(\eta^2)\,.
\\[-2.2em] \nonumber
\end{eqnarray}
The chemical
potential $\mu$ is fixed by the total particle density
$n(\mu, T) =- V^{-1}\partial \Omega (\mu, T)/ \partial \mu$. Eliminating
$\mu$ in favor of the condensate density $n_0$
via (\ref{bog4}),
we find for the particle density
\vspace{-.6em}
\begin{eqnarray}
\label{loop4}
&& n-n_0=\frac{ \eta }{V}\sum_{{\bf
k}}\frac{\epsilon({\bf k})+g_{\eff}n_0}{\sqrt{\epsilon({\bf k})^2+2g_{\eff}n_0
\epsilon({\bf k})}} \nonumber \\
&&\times \left( \frac{1}{2} +
\frac{1 }{e^{\,\beta \sqrt{\epsilon({\bf
k})^2+2g_{\eff}n_0 \epsilon({\bf k})}}-1} \right)+{\cal O}(\eta^2) \,.
\label{@nn0}
\\[-1.2em] \nonumber
\end{eqnarray}
This result is the so-called
Popov approximation \cite{And,Griffin}. It is
derived only for a small right-hand side where
$n\approx n_0$. A standard way to extend such a relation to $n \gg n_0$
is by making the equation self-consistent, replacing $n_0$ by
$n$ (and $ \eta $ by 1)
on the right-hand side.
\comment{ to obtain for the physical value $\eta =1$
\vspace{-.6em}
\begin{eqnarray}
\label{loop5}
n-n_0&=& \frac{1}{V}\sum_{{\bf
k}}\frac{\epsilon({\bf k})+g_{\eff}n}{\sqrt{\epsilon({\bf k})^2\!+\!2g_{\eff}n
\epsilon({\bf k})}}\nonumber \\&&\times
\left( \frac{1}{2} \!+\!
\frac{1}{e^{\,\beta \sqrt{\epsilon({\bf
k})^2\!+\!2g_{\eff}n \epsilon({\bf k})}}\!-\!1} \right) \, .
\\[-2.2em] \nonumber
\end{eqnarray}
}
The  location of
the quantum phase transition for all $T$
is obtained by solving this
equation for $n_0=0$
\cite{Bogoliubov,Abrikosov}. The evaluation will be discussed in
the next two paragraphs.

Note that a more systematic approach
to derive the self-consistent Popov approximation (\ref{@nn0})
proceeds by
applying variational perturbation theory
to (\ref{bog4}) and (\ref{bog6}) according to the rules
developed in
\cite{VPT1,SC}, and applied successfully
to critical phenomena in
\cite{VPT2} as well as many other strong-coupling problems
\cite{Festschrift}.
There one introduces a dummy variational parameter $\tilde \mu$
by replacing
$\mu\rightarrow \tilde \mu+ \eta r$
with
$r\equiv (\mu-\tilde \mu)/ \eta $,
and re-expands
consistently at fixed $r$ up to the first power in $ \eta $.
After this  one re-inserts   $r\equiv (\mu-\tilde \mu)/ \eta $
and extremizes the resulting expression
with respect to $\tilde \mu$.
The result turns out to be precisely the self-consistent
version of the
Popov approximation
 (\ref{@nn0})
(see also the related discussion of the self-consistent Hartree-Fock-Bogoliubov-Popov approximation
in one-dimensional optical lattices \cite{HFB1,HFB2}).\\

{\bf 3)} We first discuss the formation of a condensate for the
free-particle spectrum $\epsilon({\bf k})=\hbar ^2 {\bf k}^2/2M_{\eff}$,
 where
momentum sums
reduce to
$\sum_{{\bf k}}\rightarrow V\int {d^D k}/{(2\pi)^D}\,.$
The integral of the zero-temperature contribution can now
be evaluated analytically, and we
obtain in $D=3$ dimensions
%the following equation
for the transition curve in the $T-a_{\eff}$ plane:
\vspace{-.6em}
\begin{eqnarray}
\label{phase}
a_{\eff}\, n^{1/3}
\left[ 1 +\frac{3 \alpha}{16} I(\alpha)\right]^{2/3} = \left(\frac{9 \pi}{64} \right)^{1/3}.
\label{@EQA}
\\[-1.2em] \nonumber
\end{eqnarray}
Here $I( \alpha )$ abbreviates the integral
\vspace{-.6em}
\begin{eqnarray}
\label{I}
I(\alpha)=\int_0^\infty
dx \, \frac{x\alpha+8}{2\sqrt{x\alpha +16}
 \,\, ( e^{\, \sqrt{x^2\alpha/16 +x}}-1) }\, ,
\\[-1.2em] \nonumber
\end{eqnarray}
\begin{figure}[t]
\unitlength1mm
\begin{picture}(0,45)
\put(-40,0){\epsfxsize=6cm \epsfbox{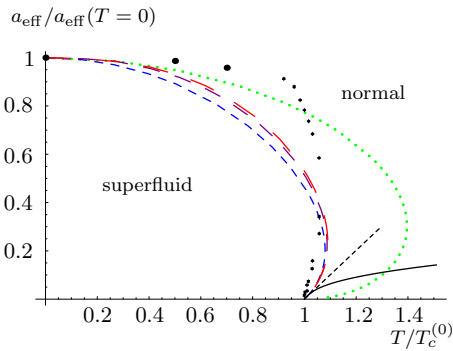}}
\end{picture}
\caption{\label{Phase1}
Quantum phase diagram
of a homogeneous dilute Bose gas in
variationally improved one-loop approximation without (dotted)
and with properly imposed  higher-loop slope properties
at $T_c^{(0)}$ (dashed length increasing with VPT order from 1 to 3).
Short solid curve starting at $T=T_c^{(0)}$ is due to a second-order finite-temperature calculation of
Arnold et al. \cite{Ar}.
Dashed straight line indicates the slope of our curve extracted either from
Monte-Carlo data \cite{Svistunov,Arnoldnum} or precise high-temperature calculations \cite{kleinertbec,Boris}.
Diamonds correspond to the Monte-Carlo data of Ref.
\cite{Ceperley} and dots stem from Ref. \cite{helium},
both scaled to their respective values $a_{\rm eff}({T=0})$.}
\end{figure}
\\[-1em]
and $\alpha\equiv {t^2}/ {a_{\eff}^{2}\, n^{2/3}\zeta(3/2)^{4/3}}$
is a dimensionless parameter,
with $t\equiv T/T_c^{(0)}$ being the reduced temperature. The
result is shown in  Fig. \ref{Phase1} a).
For small temperatures, the transition curve behaves like
\vspace{-.6em}
\begin{eqnarray}
\label{weak}
a_{\eff} n^{1/3}= a_0 + a_1 \alpha + a_2 \alpha^2 +{\cal O} \left( \alpha^3 \right) \,,
\\[-2.2em] \nonumber
\end{eqnarray}
with the dimensionless expansion coefficients
$a_0 = (9 \pi/64)^{1/3} \approx 0.762$, $a_1 = - \pi^2 a_0 /24 \approx - 0.3132$, $a_2 \approx 0.1996$,
and $a_3 \approx - 0.207$.
%The number $a_0$ already appears implicitly in the original Bogoliubov theory of superfluid helium\cite{Bogoliubov}.
The interaction causes an upward shift of the
critical temperature
from $t_c^{(0)}=1$
to
$t_c = 1 + {4\sqrt{2\pi}\sqrt{a_{\eff}n^{1/3}}}/{3\zeta(3/2)^{2/3}}
+ {\cal O} \left( a_{\eff}n^{1/3} \right)$.
This has the square-root behavior
found before in Refs. \cite{Toyoda,Huang}.

As announced in the abstract,
the phase diagram
has the interesting property
that
there exists
a {\em reentrant
transition\/} above the critical temperature
$T_c^{(0)}$
of the free system, which shows up as a {\em nose\/} in the
transition curve, where a condensate can be produced by {\em
increasing\/} $a_{\eff}$, which disappears upon
a further increase of $a_{\eff}$. Our curves agree qualitatively with early
Monte-Carlo simulations \cite{Ceperley} as shown in Fig. \ref{Phase1}.

Recent Monte-Carlo simulations \cite{Svistunov,Arnoldnum}
and precise high-temperature calculations \cite{kleinertbec,Boris} indicate,
however,
that
the square-root approximation
is unreliable near $T_c^{(0)}$, the leading critical temperature shift
being  linear in the scattering
length $a_{\eff}$ with a coefficient  $c_1 \approx 1.3$:
\vspace{-.6em}
\begin{eqnarray}
\label{right}
t_c = 1 + c_1 a_{\eff}n^{1/3}  + {\cal O} \left( a_{\eff}^2n^{2/3} \right).
\label{@new}
\\[-2.em] \nonumber
\end{eqnarray}
It is possible
to improve our self-consistent approximation
(\ref{phase}) to accommodate the high-loop result
(\ref{@new}). This can be done with
the help of variational perturbation theory
\cite{Int,VPT1,VPT2,Festschrift}. For this we use the expansion
(\ref{weak}) with a few exact coefficients and add two more trial
coefficients to enforce the behavior (\ref{@new}). This produces a
sequence of improved transition curves  shown in Fig. \ref{Phase1} a)
as dashed curves.\\
\begin{figure}[t]
\unitlength1mm
\begin{picture}(0,45)
\put(-40,0){\epsfxsize=6cm \epsfbox{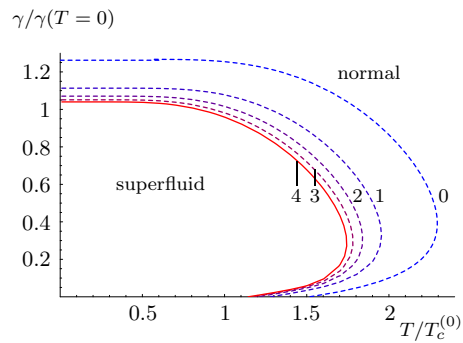}}
\end{picture}
\caption{\label{Phase2}
Quantum phase diagram of optical boson lattice
calculated for increasing hopping order (right to left).
The quantity $\gamma = U/J$
is proportional to the effective scattering length $a_{\rm eff}$.
At zero temperature, the transition takes the
superfluid into a Mott insulator.
Near $T_c$, higher corrections are needed
due to infrared singularities. Since these
are the same as in the continuum model,
the correct curve should
start out with a similar slope as in Fig.~\ref{Phase1}.  }
\end{figure}

\begin{figure}[t]
\unitlength1mm
\centerline{\includegraphics[scale=0.7]{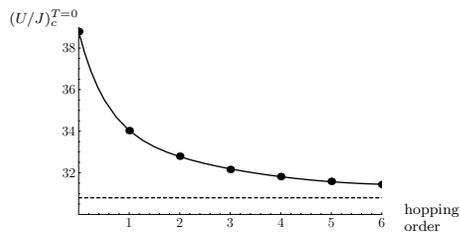}}
\caption{
Convergence of hopping expansion for
the critical value of $U/J$
at the zero-temperature
quantum phase transition. The limiting value for
$(U/J)_c^{T=0}$ is $30.8$.
}
\label{f3}
\end{figure}

{\bf 4)} We now consider an optical lattice
where the wave vectors ${\bf k}$ of the band spectrum $\epsilon_B({\bf k})$  are
restricted to the Brillouin zone
$k_i\in (-\pi/ \delta ,\pi/ \delta )$, and momentum sums
become
$%\sum_{{\bf k}}\rightarrow
V\prod_{i=1}^D
\int_{-\pi/ \deltan}^{\pi/ \deltan}{dk_i}/{2\pi}$.
The integral is
evaluated using the {\em hopping expansion\/} \cite{Kleinertdis},
in which one
expands the integrand in powers of the cosines in $\epsilon_B({\bf k})$.
By doing so, we express the result in terms of
the lattice interaction $U=g_{\eff} n$.
The transition curve is now determined by
the implicit equation
%
%\begin{eqnarray}
$F_D \left({k_B T}/{J} , {U}/{J} \right) = 0 $,
%\end{eqnarray}
%
where in zeroth hopping order
$F_D^{(0)} ( x , y ) = x -2 \sqrt{D^2 +
D y} / \ln [4 \sqrt{D^2 +  D y} + y + 2D]/[4
\sqrt{D^2 + D y} - y - 2D]$.
The resulting transition curve for $D=3$
and the next four approximations
coming from  successive
hopping orders
are shown in Fig. \ref{Phase2}. A fast convergence
is
observed, with the approximation sequence of
transition points at $T=0$ corresponding to
$(U/J)_c^{T=0}=6(3+ 2\sqrt{3})\approx 38.8,\,34.1,\,32.2,\,31.8,\,31.6\,\dots $
which converge to roughly 30.8, as shown in Fig. \ref{f3}.
Thus our value is slightly smaller than the mean-field result
$(U/J)_c^{T=0} \approx 34.8$ derived from Bose-Hubbard model
\cite{Fisher,Sheshadri,Freericks,Stoof} and the experimental
number $(U/J)_c^{T=0} \approx 36$ \cite{Greiner}. The associated
hopping sequence of transition temperatures at $U=0$ converges to
$T_c^{(0)}\approx 3.6\,J/k_B$.

For the lattice spectrum (1) we cannot improve the result near
$T_c^{(0)}$ in the same way as for the free-particle
spectrum. By analogy, we may, however, assume that the
characteristic reentrant transition will also here survive
higher-loop corrections.\\
An important question is whether the nose
survives the experimental set-up of an optical lattice where the particles require
an extra weak overall
magnetic trap
to keep them together. Indeed, if the magnetic trap is harmonic as in all present experiments
with typical frequency $~\omega _{\rm trap}\approx 2\pi\times 24$ Hz,
the nose  is expected to disappear since
by analogy with a continuum calculation,
the external trap reverses
the slope of the transition curve at $T_c^{(0)}$
\cite{arnoldtrap,YU} (see also Chapter 7 in the textbook \cite{VPT1}).
However, a nose will definitely be seen if the experiment is performed with a magnetic trap which
has an approximate box-like magnetic shape. This can be experimentally realized by using optical dipolar traps
and lasers with square profiles.
It will be interesting to study experimentally the dependence of the slope of the transition line on the shape
of the magnetic trap potential as it is distorted from harmonic to a box-like shape.
For a power-like shape of the trap \cite{Bagnato}
there should be a critical power where the nose appears \cite{Alber}.

{\it Acknowledgements:} We are very much indebted to
V.I. Yukalov for valuable
contructive criticism,
and to R. Graham and F.S. Nogueira
for further clarifying discussions.
This work was partially supported by
the ESF COSLAB Program and by the
German Research Foundation (DFG) under
Priority Program SPP 1116
and under Grant Kl-256.
One of us (S.S.)
acknowledges support from the German National Academic Foundation. 
%
%\newpage

%
\end{document}